\documentstyle[psfig]{mn}

\def\fd{\hbox{$.\!\!^{\rm d}$}}

\title[High Speed photometry of faint Cataclysmic Variables: I.]
{High Speed photometry of faint Cataclysmic Variables: I. V359 Cen, XZ Eri, HY Lup, 
V351 Pup, V630 Sgr, YY Tel, CQ Vel, CE-315}
\author[Patrick A. Woudt and Brian Warner]
       {Patrick A. Woudt\thanks{E-mail: pwoudt@artemisia.ast.uct.ac.za} 
        and Brian Warner\thanks{E-mail: warner@physci.uct.ac.za}\\
        Department of Astronomy, University of Cape Town, Private Bag,
        Rondebosch 7700, South Africa}
\date{}

\begin{document}

\maketitle

\begin{abstract}
The first results of a photometric survey of faint Cataclysmic Variables are presented. 
V359 Cen is an SU UMa star with a period of 112 min. 
Even though observed at quiescence, the mass transfer rate
in this old nova may be sufficiently high that in such a short period
system (with its implied small mass ratio) the disc may be excited into an
elliptical shape with the result that the observed brightness modulation
gives a superhump period rather than an orbital period. XZ Eri is an eclipsing dwarf nova
with an orbital period ($P_{orb}$) of 88.1 min. HY Lup has only slight variability. V351 Pup, 
the remnant of Nova Puppis 1991, has $P_{orb}$ = 2.837 h and a light curve that strongly
resembles that of the magnetic Nova Cyg 1975. V630 Sgr is the first nova remnant that 
has both positive superhumps ($P_{sh}$ = 2.980 h) and eclipses ($P_{orb}$ = 2.831 h).
The YY Tel identification is somewhat uncertain.
The correct identification for CQ Vel is provided from discovery of its
flickering activity.
The light curve of CE-315, a recently discovered AM CVn star, shows similarities to
that of GP Com, with no apparent orbital modulation.
\end{abstract}


\section{Introduction}

Cataclysmic Variable stars (CVs) are a class of interacting close binaries which attract
great attention. They enable a quantitative study of the many processes involved in 
mass transfer through accretions discs or through magnetically channeled accretion.
A list of these physical processes, which occur also in many other astrophysical 
environments, is given in Warner (2000); a review of CVs in general is given in 
Warner (1995a).

Fewer than 40 CVs have been studied in any great detail (e.g.,~with doppler tomography,
eclipse deconvolution). These, and the phenomena seen in other CVs, show that no two
CVs are exactly alike. The different combinations of masses of primaries and secondaries, 
temperatures, magnetic field strengths, orbital inclinations and mass transfer rates ($\dot{M}$) suffice to
produce great variety of behaviour. Much of this variety is already understood in
terms of the component physics accompanying the gross differences. The effects of
more subtle differences -- e.g.,~original chemical composition of the secondary, effects
on the secondary of past nova explosions, compositional and environmental (for instance,
irradiation) dependence of disc viscosity -- have still to be teased out of the 
rich observational phenomenology.

Much of the progress in the past two or three decades has come from the discovery 
of ever more variety of behaviour. There are often conspicuous effects in the longer
term light curves; examples are the ER UMa stars (dwarf novae with outbursts every
few days and superoutbursts every few weeks), `echo' outbursts after a superoutburst,
continuous small outbursts in the discs of nova-like variables. New phenomena continue
to be found also in the `close-up' details; examples are short period quasi-periodic
oscillations (from $\sim$2 s in the X-ray flux of SS Cyg in outburst to $\sim$1000 s 
modulations in many high $\dot{M}$ discs), spiral waves in discs, tilted discs,
desynchronisation of polars by nova explosions.

\begin{table*}
 \centering
  \caption{Observing log.}
  \begin{tabular}{@{}llrrrrrcl@{}}
 Object       & Type         & Run No.  & Date of obs.          & HJD of first obs. & Length    & $t_{in}$ & Tel. & $<$V$>$ \\
              &              &          & (start of night)      &  (+2451000.0)     & (h)       &     (s)   &      & (mag) \\[10pt]
{\bf V359 Cen}& SU UMa       & S6165    & 29 Dec 2000 &  908.50289  &   2.15      &      30   &  74-in & 18.8\\
              &              & S6167    & 30 Dec 2000 &  909.51368  &   1.95      &      30   &  74-in & 18.7\\
              &              & S6170    &  1 Jan 2001 &  911.51125  &   2.02      &      30   &  74-in & 18.7\\
              &              & S6186    & 20 Feb 2001 &  961.37008  &   2.05      &  30, 60   &  40-in & 18.6\\
              &              & S6189    & 21 Feb 2001 &  962.41615  &   4.03      &      60   &  40-in & 18.7\\
              &              & S6197    & 25 Feb 2001 &  966.35024  &   2.22      &      60   &  40-in & 18.6\\
              &              & S6216    & 18 May 2001 & 1048.22183  &   4.02      &      60   &  40-in & 18.6\\[5pt]
{\bf XZ Eri}  & DN           & S6156    & 26 Dec 2000 &  905.29165  &   4.02      &  20, 60   &  74-in & 19.1\\
              &              & S6158    & 27 Dec 2000 &  906.29347  &   0.36      &      30   &  74-in & 19.2\\
              &              & S6163    & 29 Dec 2000 &  908.30081  &   0.65      &  20, 30   &  74-in & 19.1\\[5pt]
{\bf HY Lup}  & NR           & S6191    & 22 Feb 2001 &  963.47488  &   3.76      &      45   &  40-in & 19.6\\[5pt]
{\bf V351 Pup}& NR           & S6157    & 26 Dec 2000 &  905.47151  &   3.06      &      15   &  74-in & 18.9\\
              &              & S6160    & 27 Dec 2000 &  906.45415  &   3.62      &      20   &  74-in & 19.1\\
              &              & S6162    & 28 Dec 2000 &  907.43106  &   3.96      &      20   &  74-in & 18.9\\
              &              & S6185    & 20 Feb 2000 &  961.27432  &   1.93      &      60   &  40-in & 18.8\\
              &              & S6192    & 23 Feb 2000 &  964.26594  &   1.20      &      60   &  40-in & 19.0\\[5pt]
{\bf V630 Sgr}& NR           & S6124    & 24 Aug 2000 &  781.22202  &   6.72      &      10   &  74-in & 17.6\\
              &              & S6126    & 25 Aug 2000 &  782.21315  &   7.15      &       5   &  74-in & 17.5\\
              &              & S6128    & 26 Aug 2000 &  783.21314  &   7.16      &   5, 10   &  74-in & 17.8\\
              &              & S6130    & 27 Aug 2000 &  784.20912  &   6.15      &   5, 10   &  74-in & 17.4\\[5pt]
{\bf YY Tel}  & SU UMa       & S6100    &  5 Jun 2000 &  701.36536  &   1.39      &      10   &  74-in & 18.6\\
              &              & S6218    & 18 May 2001 & 1048.61265  &   1.72      &  30, 60   &  40-in & 18.6\\
              &              & S6223    & 21 May 2001 & 1051.61806  &   1.57      &      60   &  40-in & 18.5\\
              &              & S6231    & 26 May 2001 & 1056.64150  &   1.07      &  30, 60   &  74-in & 18.6\\[5pt]
{\bf CQ Vel}  & NR           & S6071    & 10 Mar 2000 &  614.26263  &   4.07      &      60   &  74-in & 21.1\\[5pt]
{\bf CE-315}  & AM CVn       & S6209    & 15 May 2001 & 1045.35878  &   4.26      &      10   &  40-in & 17.6:\\
              &              & S6212    & 16 May 2001 & 1046.25888  &   6.63      &      20   &  40-in & 17.6:\\[10pt]
\end{tabular}
{\footnotesize 
\newline 
Notes: NR = Nova Remnant, DN = Dwarf Nova, $t_{in}$ is the integration time.\hfill}
\label{tab1}
\end{table*}

Although some of the new phenomena have accompanied the discovery of new CVs  (e.g., 
the $\sim 10^8$ G magnetic fields in AR UMa and V884 Her), most have been discovered 
in long known CVs (e.g., TeV gamma rays from AE Aqr, ZZ Cet pulsations in GW Lib).
It is clear, therefore, that while the great majority of CVs remain under-observed
new phenomena may remain undiscovered. At a less exotic level, the determination of
a significant number of new orbital periods, and in particular the discovery of eclipsing
systems, provide objects for more detailed structural studies. With the advent of 8-m class
telescopes such studies can be made even to 20th magnitude.
The ongoing survey, of which this paper presents the first general results, was 
initiated with this in mind.

The need for such a survey, particularly in the southern hemisphere and capable of
succeeding in the crowded fields in the general direction of the Galactic centre where the
majority of nova remnants reside, was recognised in the early 1990s and as a result
the CV group at the University of Cape Town (UCT) used several years' equipment funds (in advance)
for the purchase of a state-of-the-art CCD detector. The photometer and its software
were designed and constructed by Dr.~D.~O'Donoghue, who was a member of the Astronomy 
Department at UCT at that time. The UCT CCD photometer, completed in 1993, has been
extensively used on telescopes at the Sutherland site of the South African Astronomical
Observatory. An outline of its functional properties can be found in O'Donoghue (1995) and
Koen \& O'Donoghue (1995). 
In the CV group at UCT it was first largely used for observations of well known CVs at
higher time resolution or improved signal-to-noise. Its use for a general survey of 
faint CVs started in 1997, with the almost immediate discovery of the first ZZ Cet (i.e., 
non-radial) pulsations in the white dwarf primary of the SU UMa type dwarf nova GW Lib 
(Warner \& van Zyl 1998) and independent discovery of the intermediate polar nature of
the nova remnant HZ Pup (see Abbott \& Shafter 1997 for detailed description). Follow-up
observations of these led to neglect of the survey for some time, but we are now actively
observing faint CVs again. We present here initial results.

The observations are given in Section 2 and some brief conclusions in Section 3.

\section{Observations}

All observations have been made with the UCT CCD photometer attached either to
the 40-in or 74-in reflectors at Sutherland. In order to capture as many photons 
as possible `white light' was used, with the result that extinction corrections
and magnitude calibrations are only approximate.

The observing log for the stars included in this paper is given in Table 1.

\subsection{V359 Centauri}

V359 Cen was discovered bright in April 1929\footnote{Note that Duerbeck (1987) states 1939, and this
has been propagated in later literature.} and entered the lists as a probable nova, unconfirmed because
no spectra were obtained. There are still no spectra available at minimum light, but in recent years
further outbursts to a maximum of V $\sim 13.8$ have been observed at intervals $\sim 1$ y, showing the
star to be of SU UMa type. The quiescent magnitude given in the Downes, Webbink \& Shara~(1997) catalogue is 21.0
from J plates, and Munari \& Zwitter (1998) gave V $\ge 20.5$ in 1996, but we find V $\sim$ 18.7, which
is a great deal easier to observe at moderate time resolution. Our May 2001 observation was obtained
three weeks after a superoutburst, but the star had returned to its quiescent brightness.

\begin{figure}
\centerline{\hbox{\psfig{figure=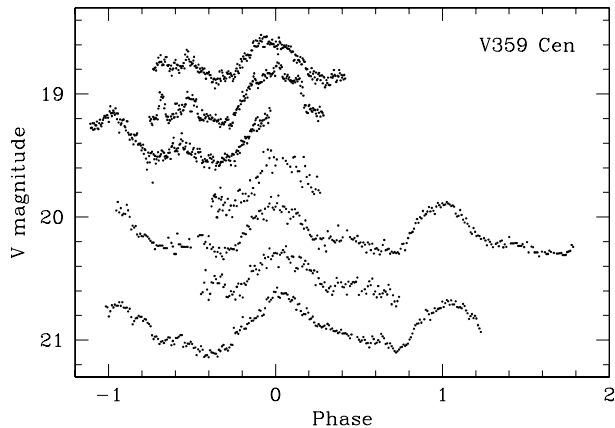,width=8.2cm}}}
  \caption{Light curves of V359 Cen at quiescence phased according to the 112 min period. The light
curves are plotted top to bottom in the same order as in Table 1. All light curves, except for the top one,
are displaced along the magnitude axis for display purposes only. The mean magnitude of V359 Cen for each run is given in Table 1.}
 \label{lcv359cen}
\end{figure}

Our light curves are displayed in Fig.~\ref{lcv359cen} and show a recurrent hump with an amplitude
$\sim 0.4-0.5$ mag. Our observations are not distributed adequately to remove aliases; the best
determined period from the February 2001 run is 112 min (0.0779 days). The variations of hump 
profile and their occasional almost triangular shape resemble more what is seen in superhumps than
orbital modulations, but it is rare to observe superhumps at quiescence -- the only well documented
examples are EG Cnc which sustained `late' superhumps for at least 90 d after the 1996 superoutburst (at
a time when `echo' normal outbursts were occuring) (Patterson et al.~1998) and AL Com with a persistent
superhump as well as an orbital hump at quiescence (Patterson et al.~1996). If what we are seeing
is an orbital hump then the inclination of V359 Cen must be $\sim 65^\circ$.

\subsection{XZ Eridanus}

XZ Eri has been known as a probable dwarf nova for nearly 70 years (Shapley \& Hughes 1934). 
In recent years only three outbursts
have been detected: Mar/Apr 1995, Jan 1998, Feb 1999. The first of these reached V $\sim 14.1$, the others
reached V $\sim 14.6$. CCD photometry at quiescence gives V = 18.7 (Howell et al.~1991). The rarity and
range of the outbursts suggests that they are superoutbursts. Any normal outbursts may have gone 
unnoticed. There has been no high speed photometry during an outburst.

A quiescence spectrum, obtained by Szkody \& Howell (1992) shows the Balmer emission and continuum
characteristic of a dwarf nova. There is a hint of doubling in the strongest Balmer lines, indicative
of a high inclination system. Nearly 4 h of CCD photometry through a $V$ filter at quiescence, with 220 s
integration on a 1.8-m telescope, showed a scatter of $\sim 1.0$ mag and, to quote the authors, no
indications of periodic modulation (Howell et al.~1991).

\begin{figure}
\centerline{\hbox{\psfig{figure=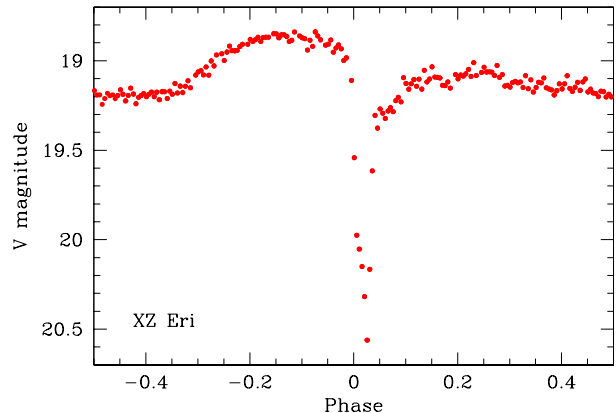,width=8.2cm}}}
  \caption{The mean light curve of XZ Eri.}
 \label{lcxzeri}
\end{figure}

Our observations (Fig.~\ref{lcxzeri}) show XZ Eri to be, in brief, a Z Cha lookalike, with the succession
of immersions and emersions of the white dwarf and bright spot all quite clear. At V $\sim 19$
we cannot achieve sufficient time resolution to separate these components as clearly as in Z Cha (e.g., Wood
et al.~1986). Using our best estimate for the time of mid-eclipse of the white dwarf we derive
the following ephemeris:

\begin{equation}
{\rm HJD_{min}} = 2451905.4419 + 0\fd0612 \, {\rm E}.
\label{eph359}
\end{equation}

The short $P_{orb}$ (1.47 h = 88.1 min) implies that XZ Eri will have superhumps 
when observed in outburst.

\subsection{HY Lupi}

HY Lup was Nova Lupi 1993, which reached V $\sim 8$ and was moderately fast ($t_2 \sim 15$ d). It is 
currently at V $\sim 18.9$ and has a resolved ejecta shell (Downes \& Duerbeck 2000). No pre-outburst
candidate brighter than V $\sim 17$ is detectable (McNaught \& Garradd 1993) and it is 
probable that the remnant has not yet reached its quiescent brightness. No high speed photometry
has been published for this nova remnant. 

\begin{figure}
\centerline{\hbox{\psfig{figure=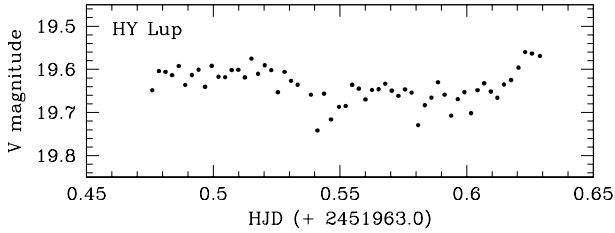,width=8.2cm}}}
  \caption{The light curve of HY Lup obtained on 22 February 2001. The data
are binned along the horizontal axis by a factor of five.}
 \label{lchylup}
\end{figure}

Our photometry (Fig.~\ref{lchylup}) shows little evidence for rapid flickering and only slow
variations of small amplitude. There is no suggestion of any orbital modulation which would encourage
further photometry. The $P_{orb}$ for HY Lup will need to be obtained spectroscopically.

\subsection{V351 Puppis}

Nova Puppis 1991 (V351 Puppis) was a moderately fast nova ($t_2 = 16$ d), probably a neon nova
(Pachoulakis \& Saizar 1995). The pre-eruption magnitude was variable over the range 
$21 - > 22$ (McNaught 1992). The eruption light curve through 1996 is illustrated in Saizar 
et al.~(1996); in 1998 the remnant was at V $\sim 19.6$ (Downes \& Duerbeck 2000).

There are one or two additional aspects of this nova that are of interest. Sixteen 
months after optical maximum, hard X-rays were detected by ROSAT which, from the absence
of a soft X-ray component, were deduced to be caused either by shocks in the ejected
shell or by magnetically controlled accretion onto the white dwarf (Orio 
et al.~1996). Variability (time scale $<$ days) observed in the X-ray flux seemed to
favour the latter interpretation. However, a later analysis explains the absence of
soft X-rays as due to internal and external absorption, allowing the hard
X-rays to be the product of surface nuclear burning, as in most novae (Vanlandingham et al.~2001).
Multiwavelength observations of the nebular phase of the eruption made in 1992-1993 revealed 
a strong red excess attributed to the secondary star, implying that this component is
either a giant or an irradiated dwarf (Saizar et al.~1996). 
In the former case an orbital period $P_{orb} \sim$ days would be expected, in the
latter $P_{orb} \la 4$ h would be required in order for the secondary to be close
enough to the hot primary to be strongly irradiated.
The relatively small mass of the ejected shell of N Puppis 1991 implies a high 
mass for the white dwarf -- near the Chandrasekhar limit (Vanlandingham et al.~2001).

\begin{figure}
\centerline{\hbox{\psfig{figure=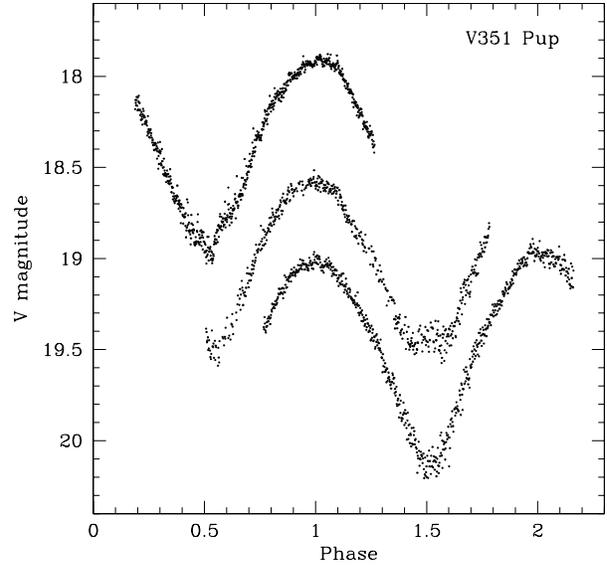,width=8.2cm}}}
  \caption{The light curves of V351 Pup obtained in December 2000, phased 
according to Eqn.~\ref{eph351}. A vertical shift of 0.5 mag has been introduced between light curves
for display purposes.}
 \label{lcv351pup}
\end{figure}

We observed V351 Pup in December 2000 and found it to be modulated in brightness with a range
of up to $\sim$1.2 mag and a period of 2.837 h. The light curves are given in Fig.~\ref{lcv351pup}
and show variable amplitude, low amplitude flickering, rounded maxima and a V-shaped
minimum which may be filled in to form a flattened region. The ephemeris from these
observations is

\begin{equation}
{\rm HJD_{max}} = 2451905.56746 + 0\fd1182 \, {\rm E}.
\label{eph351}
\end{equation}

The additional observations we made in February 2001 are too far removed in time
to enable us to improve on the precision of the period.

Our light curves for V351 Pup have a striking similarity to those of V1500 Cyg in the late
1980s (see especially Fig.~1 of Kaluzny \& Chlebowski 1988). V1500 Cyg was Nova Cygni 1975 and in the 
late 1980s it was about 15.3 mag below its maximum brightness. Nova Pup 1991 was not discovered until after 
maximum, but there are reasons to believe it did not exceed V $\sim 5.5$ (Saizar et al.~1996) and it therefore
at present is $\sim$14 mag below maximum. V1500 Cyg is a desynchronised polar (Stockman, Schmidt \&
Lamb 1988; Kaluzny \& Chlebowski 1988) with an orbital period of 3.35 h, which suggests the possibility 
that V351 Pup may also be magnetic. With this in mind we asked Gary Schmidt to check for
circular polarization and he reports ``A sequence of spectropolarimetric observations 
was obtained by G.~and P.~Schmidt on 2 January 2001 using the 74-in reflector at Mt.~Stromlo and covering
the range 4220 -- 7300 {\AA} at $\sim$10 {\AA} resolution. The summed spectrum, which spans a total of 3.7 h, 
reveals a weak continuum (V $\sim 19.4$) that is approximately flat in $f_{\lambda}$, plus broad
($\pm 1700$ km s$^{-1}$) nebular emission lines of H, He\,II and O\,III. The polarimetric results 
fail to show significant circular polarization anywhere in the observational sequence, and the
coadded sum is unpolarized to a 3$\sigma$ upper limit of 1\%''.

Before dismissing the possibility that V351 Pup is a magnetic nova we note that (a) the level of circular 
polarization in V1500 Cyg is $\sim 1.5$\% (Stockman et al.~1988), which, compared with
the 10--30\% in polars, shows the presence of much unpolarized light, probably that from
irradiation of the secondary by the still hot primary; (b) V351 Pup is 10 y from its eruption
and less far in its recovery than was V1500 Cyg in the late 1980s -- and should therefore
have even more unpolarised light; (c) GQ Mus (Nova Mus 1983) also has a polar-like 
orbital light curve and shows strong spectral signatures of magnetic accretion, but no
circular polarization (Diaz \& Steiner 1994); (d) similar remarks apply to V2214 Oph (Nova
Oph 1988: Baptista et al.~1993).

\subsection{V630 Saggitarii}

\begin{figure*}
\centerline{\hbox{\psfig{figure=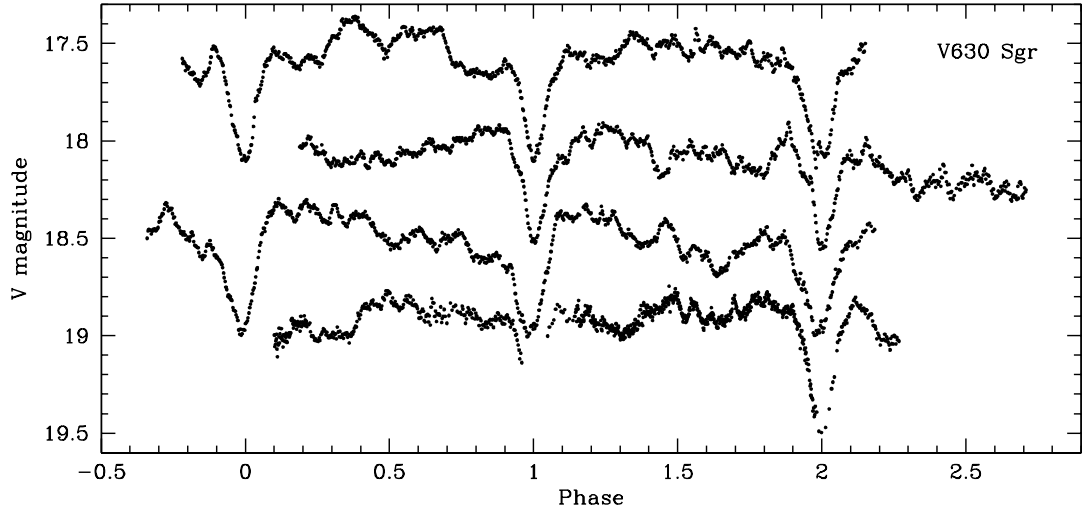,width=16cm}}}
  \caption{The light curves of V630 Sgr obtained in August 2000, phased 
according to Eqn.~\ref{eph630}. These have been binned to 30 s to give greater 
clarity of reproduction. The light curves are plotted top to bottom in the same
order as in Table 1. All light curves, except for the top one, are displaced
along the magnitude axis for display purposes only. The mean magnitude of V630 Sgr
for each run is given in Table 1.}
 \label{lcv630sgr}
\end{figure*}

V630 Sgr was the last of four novae detected in Saggitarius in 1936. It was discovered 
in October of that year, at a time when Harvard objective prism plates were being exposed,
whereby it attained the status and alias HD 321353. The observed maximum was V $\sim 4.5$, but it
probably reached V $\sim 1.6$ (Downes et al.~1997), and at minimum it is V $\sim 19$, giving
it the unusually large eruption amplitude of $\sim 17.5$ mag, even for a $t_2 = 4$ d nova, suggesting
a high inclination system (Warner 1986: this conclusion arises from the apparent intrinsic faintness
of the nearly edge-on accretion disc which such an object has in quiescence).

There have been no modern photometric or spectroscopic observations of V630 Sgr, which is a faint
object in a very crowded field. It is listed by Diaz \& Steiner (1991), on the basis
of its similarity to V1500 Cyg in being a very fast nova of large amplitude, as a candidate
magnetic system.

Our light curves are shown in Fig.~\ref{lcv630sgr} and illustrate that V630 Sgr is indeed a moderately high 
inclination system, with eclipses 0.4 -- 0.6 mag deep. The ephemeris derived from our light curves,
measured from the fundamental Fourier component. The time of the first minimum of our first light curve
of this set is given in the ephemeris, which is

\begin{equation}
{\rm HJD_{min}} = 2451781.2480 + 0\fd1180 \, {\rm E}.
\label{eph630}
\end{equation}

The orbital period of 2.831 h is just at the top of the orbital period gap.

\begin{figure}
\centerline{\hbox{\psfig{figure=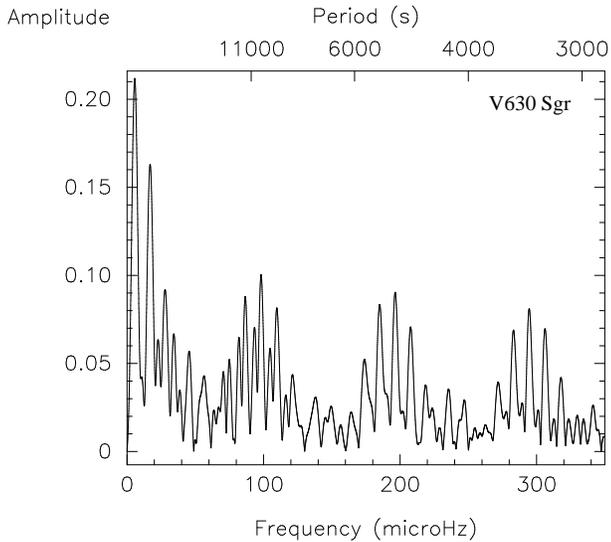,width=8.2cm}}}
  \caption{The Fourier transform of the light curves shown in Fig.~\ref{lcv630sgr}.}
 \label{fsv630sgr}
\end{figure}

\begin{figure*}
\centerline{\hbox{\psfig{figure=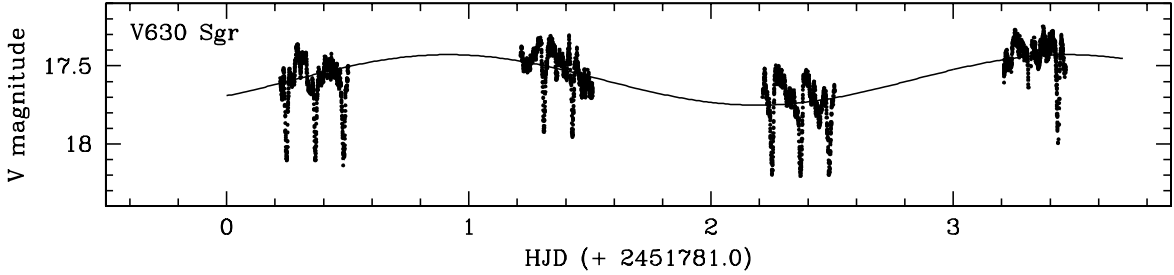,width=16.0cm}}}
  \caption{The long term optical behaviour of V630 Sgr.}
 \label{ltv630sgr}
\end{figure*}

It is evident from Fig.~\ref{lcv630sgr} that there is a hump of complex and variable profile
that has a period comparable to $P_{orb}$. The Fourier amplitude spectrum of the complete set 
of light curves is shown in Fig.~\ref{fsv630sgr}. The main groups of power are associated
with the fundamental and strong harmonics of the narrow eclipses. The window pattern of the data set
can be judged from the first and second harmonics.  Superimposed, but slightly offset from
the $P_{orb}$ fundamental window pattern, is a second pattern. Prewhitening at the orbital
frequency gives a clear pattern with an amplitude of 0.07 mag centred on $P_sh = 2.980$ h, which is the period of the drifting
humps. Therefore, V630 Sgr is {\sl a nova remnant that is also an eclipsing permanent superhumper}. 
This is the first CV found to combine all of these features. The superhump excess of 5.3\%
and beat period of 2.4 d are typical for superhumps in superoutbursting dwarf novae and
novalike variables (Warner 1995a). The beat period is evident in the night-to-night
variation of mean brightness, as seen in Fig.~\ref{ltv630sgr}. The Fourier transform of this delivers 
an approximate period of 2.1 d and an amplitude of 0.22 mag.

The eclipse depth of $\sim 0.6$ mag is relatively shallow, but at the
$P_{orb}$ of V630 Sgr it implies a quite high orbital inclination. The width at half depth
of the narrowest eclipses is $\Delta \psi \sim 0.085 P_{orb}$. The secondary mass $M(2)$ at
$P_{orb} = 2.8$ h is $\sim 0.24$ M$_\odot$ (Warner 1995a) and as V630 Sgr 
was a very fast nova we expect $M(1) \sim 1.2$ M$_\odot$, giving a mass ratio $q \sim 0.20$. 
These figures are not quite compatible with the theoretical $\Delta \psi, q, i$ relationship
(Horne 1985), but they serve to show that $i \ge 85^\circ$. The presence of superhumps 
shows V630 Sgr to be a high $\dot{M}$ system which will have a disc radius at the 3:1 resonant radius,
which is $r_d = 0.45 a$. The Roche lobe radius for the secondary is $R(2) = 0.25 a$ for $q = 0.2$. Therefore,
even at large inclination, the eclipses are partial, as is seen in the eclipse profiles
in Fig.~\ref{lcv630sgr}.

The observed eruption range of 17.5 mag is 3.5 mag greater than expected for a $t_2 = 4$ d nova
(see Fig.~5.4 of Warner 1995a). Using the correction formula (Eqn.~2.63 of Warner 1995a) for disc
inclinations, removes the 3.5 mag excess if $i \sim 88^\circ$. We conclude that V630 Sgr is a very
high inclination system.

\subsection{YY Telescopium}

YY Tel is a dwarf nova with a range V $\sim 14.4-19.3$ (Downes et al.~1997)
suspected to be an SU UMa star of long outburst interval (O'Donoghue et al.~1991). The 
last reported outburst was in October 1998, and no high speed photometry has been reported, either
in outburst or at quiescence. The candidate identified at quiescence (Downes et al.~1997) was
observed spectroscopically by Zwitter \& Munari (1996) and showed only a continuum with
no emission lines characteristic of a CV. Their photometry gave V = 19.26, (B-V) = 0.47, 
(U-B) = -0.28, placing the candidate near the black body line at $T \sim 6700$ K in
the two colour diagram. This is also a region frequented by late F type subdwarfs.

\begin{figure}
\centerline{\hbox{\psfig{figure=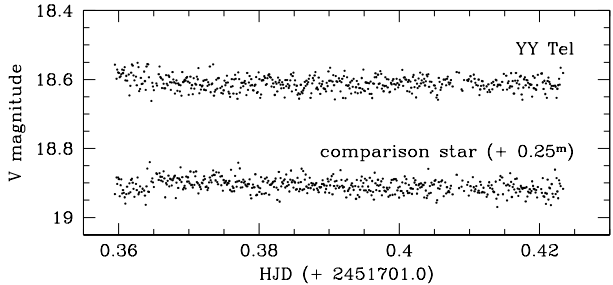,width=8.2cm}}}
  \caption{The light curve of YY Tel obtained on 6 June 2000. The lower light curve
shows a comparison star of the same brightness (displaced by 0.25 mag for display purposes).}
 \label{lcyytel}
\end{figure}

\begin{figure}
\centerline{\hbox{\psfig{figure=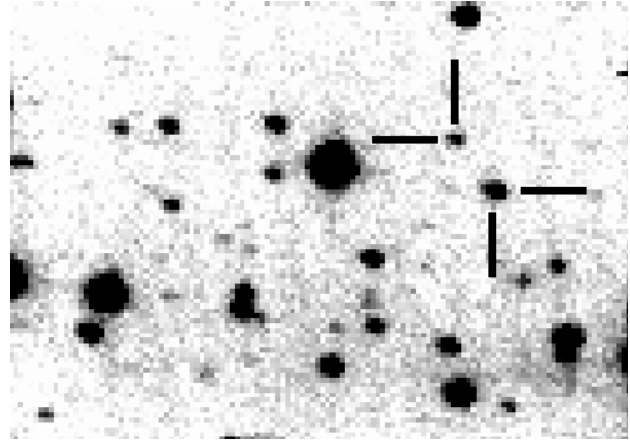,width=8.2cm}}}
  \caption{CCD image of YY Tel (the brighter star of the two indicated by markers) taken 
on 21 May 2001. The field of view is 109$''$ by 74$''$, north is up and east is to the left.}
 \label{fcyytel}
\end{figure}

Our observation on 5 June 2000 showed the YY Tel candidate to be constant in brightness (see Fig.~\ref{lcyytel}). Re-examination
nearly one year later again showed absence of any rapid variation but the star changed in brightness
by 0.15 mag over three to five days. A nearby, fainter star, marked in Fig.~\ref{fcyytel} (YY Tel is the brighter of the
two stars that are marked), shows marginal evidence for variation and should be examined spectroscopically.

\subsection{CQ Velorum}

\begin{figure}
\centerline{\hbox{\psfig{figure=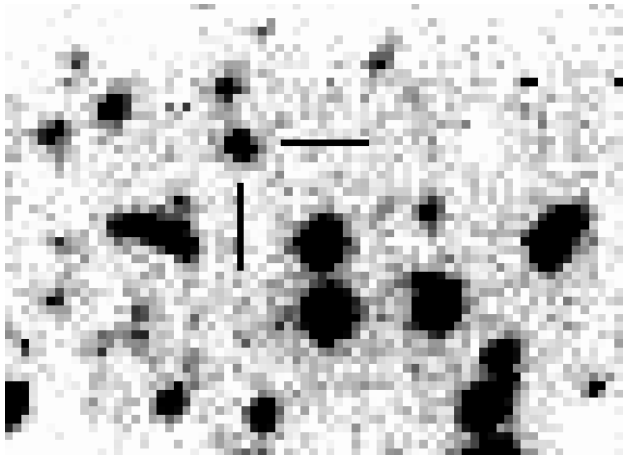,width=8.2cm}}}
  \caption{CCD image of CQ Vel (indicated by the markers) taken on 10 March 2000.
The field of view is 50$''$ by 34$''$, north is up and east is to the left.}
 \label{cqvelfc}
\end{figure}

\begin{figure}
\centerline{\hbox{\psfig{figure=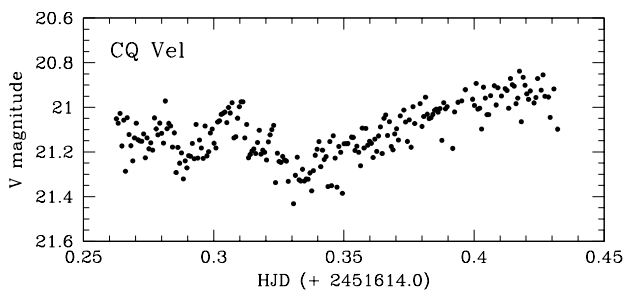,width=8.2cm}}}
  \caption{The light curve of CQ Vel (V = 21.1) obtained on 10 March 2000. }
 \label{lccqvel}
\end{figure}

CQ Vel was Nova Velorum 1940, reaching V $\sim 9.0$.  It has been identified with a 21st
magnitude star (Duerbeck 1987) but no confirmatory spectra or photometry have been obtained. In our
observations on 10 March 2000 we found the candidate nova remnant to be of constant brightness as
were all the stars in the CCD frame, except for a star slightly fainter than the candidate, 9$''$ to the 
northwest (Fig.~\ref{cqvelfc}). This star has the flickering characteristic of a CV, see Fig.~\ref{lccqvel}.
These observations provide a correct identification for CQ Vel.

\subsection{CE-315}

CE-315 is a recently discovered member of the AM CVn class of CVs: systems transferring
helium rather than hydrogen (Ruiz et al.~2001). It is only the second of this class to be
found in a long-term state of low $\dot{M}$, the other being GP Com (alias G61-29). Ruiz et al.~found
a period of 65.1 $\pm$ 0.7 min from an S-wave in the spectra, which by analogy with GP Com is thought
to be the orbital period.

\begin{figure}
\centerline{\hbox{\psfig{figure=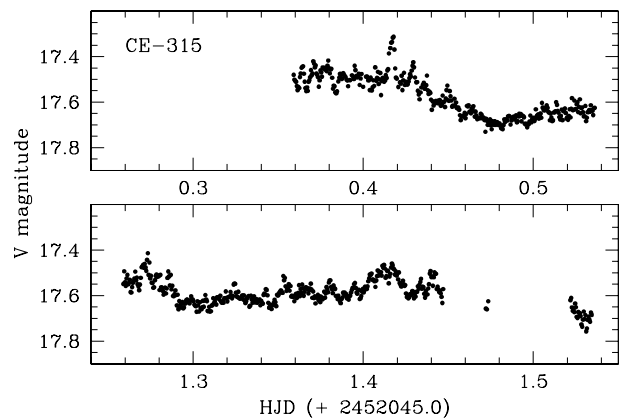,width=8.2cm}}}
  \caption{The light curves of CE-315 obtained on 15 and 16 May 2001. The data in the top panel have
been binned along the horizontal axis by a factor of four, the data in the bottom panel by a factor
of two. }
 \label{lcce315}
\end{figure}

We have obtained two light curves for CE-315, which are displayed in Fig.~\ref{lcce315}. There is no
sign in the light curve, or its Fourier Transform, of the orbital periodicity. CE-315 resembles GP Com 
in the general appearance of its light curve (see Warner 1972; Harrop-Allin 1996): much short time scale
flaring or flickering set on a slowly varying background. In GP Com there are occasions when a few cycles
at the 46.5 min orbital period appear in the light curve (Harrop-Allin 1996); if the same occurs in CE-315
it would need extensive photometry to detect.

The fact that the two longest period AM CVn stars, GP Com and CE-315, are low $\dot{M}$ systems
(equivalent to dwarf novae in hydrogen transferring CVs) is in accord with the evolutionary model where
orbital angular momentum is the result of radiation of gravitational waves (Warner 1995a,b).
According to Eqn.~9.61 of Warner (1995a) the mass transfer rates should be $\dot{M} \sim 1.0\times 10^{-11}$
M$_{\odot}$ y$^{-1}$ for GP Com and $\dot{M} \sim 3\times 10^{-12}$ M$_{\odot}$ y$^{-1}$ for CE-315. GP Com has X-rays 
modulated at the orbital period (Beuermann \& Thomas 1993). At B $\sim 17.2$, CE-315 is $\sim 1.4$ mag fainter
than GP Com, but this is accounted for by the lower $\dot{M}$, so they are at similar distances
and it would be worth seeking an X-ray detection.

\section{Conclusions}

The results presented here cover a miscellany of CV types and demonstrate the rewards waiting
among the fainter members of this class. For those objects where new phenomena have been found
(e.g., the possible quiescent superhumps in V359 Cen), or where rarity is an incentive (e.g., the
eclipsing superhumping nova remnant V630 Sgr), further observations with larger telescopes are 
required. V351 Pup may show circular polarization or other magnetic signatures when it has descended
the nova light curve further.

\section*{Acknowledgments}
We thank Dr. D.~O'Donoghue for the use of his EAGLE program for Fourier analysis of the light curves. PAW is 
funded partly through strategic funds made available to BW by the University of Cape Town. BW's research
is funded entirely by that university.

\end{document}